\documentstyle[12pt]{article}
\date{}
\def\baselinestretch{1.3}
\begin{document}

\begin{center}
{\large \bf On the Forward Scattering Amplitude \\
of the Virtual Longitudinal Photon \\
at Zero Energy.}

\vspace{3mm}
B.L.Ioffe\\
{\small \it Institute of Theoretical
and Experimental Physics}\\
{\small \it B.Cheremushkinskaya 25, 117259, Moscow, Russia}

\vspace{5mm}
\small
\parbox{14cm}{The theorem is proved, determining the forward scattering
amplitude of virtual longitudinal photon at zero energy on any hadronic
target in the limit of small photon virtualities $Q^2$. The theorem is
strict, based only on Lorenz and gauge invariance.  No assumptions about the
strong interaction, besides the gap in the mass spectrum, are used. Two
terms in the expansion over $Q^2$ are calculated. The estimation  is given
up to what $Q^2$ the elastic term can represent  the whole amplitude of
longitudinal virtual photon-proton scattering.}

\vspace{5mm}
\noindent
\parbox{14cm}{PACS numbers: 03.80.+r,  11.55.-m, 11.55.Fv., 13.60.-r}

\end{center}
\newcommand{\be}{\begin{equation}}
\newcommand{\ee}{\end{equation}}

\def\la{\mathrel{\mathpalette\fun <}}
\def\ga{\mathrel{\mathpalette\fun >}}
\def\fun#1#2{\lower3.6pt\vbox{\baselineskip0pt\lineskip.9pt
\ialign{$\mathsurround=0pt#1\hfil##\hfil$\crcr#2\crcr\sim\crcr}}}

\def\baselinestretch{1.5}
\normalsize

\vspace{2mm}
The knowledge of the forward scattering amplitude of the virtual
longitudinal photon is desirable for many purposes. This amplitude appears
in the electron-nucleon scattering, in the approaches, connected with
vector dominance model etc.  The forward scattering
amplitude at zero photon energy arises as a subtraction term in the
dispersion relation for this amplitude.

In this paper I prove the theorem, determining the value of this amplitude
at zero photon energy at small photon virtualities. This theorem is similar
to the famous theorem, proved many years ago by Thirring [1], Kroll
and Ruderman [2], Low [3]  and Gell-Mann and Goldberger [4],
where it was shown that in the case of real (i.e.transverse) photon the
scattering amplitude at zero photon energy on any target is given by Thomson
formula:  $f_T(0)=-Z^2\alpha/m$ [1,2] and the terms linear in the
photon energy for spin $1/2$  target are expressed through the target static
magnetic moment [3,4]. (Here $\alpha$ is the fine structure constant,
$Z$ and $m$ are the target charge and mass.)

The virtual photon forward scattering amplitude $f(\nu,q^2)$ is the function
of two invariants $\nu=pq$ and $q^2$, where $p$  and $q$  are the target and
photon 4-momenta. I am interested in the case $\nu=0$  and $q^2<0$
(metrics: $q^2 = q^2_0 - {\bf q}^2$). It is convenient to go to the Lorenz
coordinate system, where $q_0=0$ and the $z$-axis is the collision axis. In
this coordinate system $q^2=-q^2_z$  and the condition $\nu=0$ means
that $p_z=0$ -- the coordinate system coincides with the laboratory system.
The forward scattering amplitude of longitudinal virtual photon is given by
\be
f_L(\nu,q^2) = e_{L\mu}T_{\mu\lambda}(\nu,q^2)e_{L\lambda},
\ee
where $e_{L\mu}$ is the photon longitudinal polarization,
\be
T_{\mu\lambda}(\nu,q^2) = ie^2\int d^4x e^{iqx}\langle
p\mid T\left \{
j_{\mu}(x),~j_{\lambda}(0)\right\} \mid p \rangle,
\ee
and $j_{\mu}(x)$ is the electromagnetic current. From the condition $e_Lq=0$
in the chosen coordinate system we have $e_0=1, {\bf e}=0$. Therefore
in this coordinate system
\be
f_L(\nu,q^2) = ie^2 \int d^4 x e^{-iq_zz}\langle
p\mid T\left \{
j_0(x),~j_0(0)\right\} \mid p \rangle,
\ee
Represent (3) as the sum over the whole set of intermediate states and
integrate over the time, using

$$j_0(t,{\bf x}) = e^{iHt} j_0(0,{\bf x})e^{-iHt},$$
where $H$ is the Hamiltonian. At $\nu=0$  we have
$$f_L(0,q^2) = -e^2 \sum_n \int d^3 x e^{-iq_z z}\frac{1}{E_p-E_n} \Biggl [
\langle p \mid j_0(0,{\bf x})\mid n \rangle \langle n \mid j_0(0)\mid p
\rangle $$

\be
+\langle p \mid j_0(0)\mid n \rangle \langle n \mid j_0(0,{\bf x})\mid p
\rangle \Biggr ],
\ee
where $E_p=m$ and $E_n$ are the energies of initial and intermediate states,
$m$ is the target mass. Go to the limit $q_z \to 0$ and integrate over $d^3x$
in (4). Then
\be
\int d^3x ~j_0(0, {\bf x}) = {\bf Z}
\ee
-- the total charge operator. Among the intermediate states in (4) only
the target state gives nonvanishing contribution to (4). Its energy is equal
to
\be
E_n = \sqrt{m^2+q^2_z} \approx m + \frac{1}{2}\frac{q^2_z}{m}
\ee
From (4),(5),(6) we get the general formula $(Q^2=-q^2 > 0)$
\be
f_L(0,Q^2) = 4\alpha Z^2 \frac{m}{Q^2},
\ee
where $Z$ is the target charge. Remarkable features of eq.7 are  the
singularity in $Q^2$  and the sign opposite to the Thomson amplitude.

The
derivation of eq.7, is based only on  charge conservation. It holds for any
strong interacting target and no assumptions about strong  interaction,
besides the gap in the mass spectrum are used. The latter is necessary,
because, if the intermediate states would be degenerate, the proof, in
general, would fail. For this reason the case of QED, where such
degeneration takes place due to infrared problem, requires a special
investigation.

The proof is valid strictly at $\nu=0$.
At $\nu\not=0$ and $Q^2 \to 0~ f_L(\nu,Q^2) \sim Q^2$, as could be expected
{\it a priori}. (The formal reason, why the presented above derivation fails
at $\nu\not=0$, is that in the chosen coordinate system, the initial state
momentum $p_z=-\nu/q_z$ and goes to infinity at $Q^2\to 0$.). However, the
proof may be generalized to the "scaling case" $\nu \sim Q^2 \to 0$ (see
below, eq.10).

For spin zero target the next order term in $Q^2$ can also  be calculated
from (4). Again, only intermediate state, coinciding with the initial one,
gives a nonvanishing contribution in this order. The second order term in
$Q^2$ arises from two sources: 1) from the expansion of the exponent in (4)
up to the second order; 2) from the next term in the expansion of the
denominator in $q^2_z$. As a result we get instead of (7) the following
formula, correct up to $O(Q^2)$ terms:
\be
f_L(0,Q^2)_{spin~0} = 4Z^2\alpha \frac{m}{Q^2}\Biggl [ 1 + \frac{Q^2}{4m^2}
- \frac{1}{3} Q^2 \langle r^2_E \rangle \Biggr ],
\ee
where $\langle r^2_E \rangle$ is the target mean square charge radius
\be
\langle r^2_E \rangle = \langle p \mid \int d^3 x {\bf r}^2 j_0(x)\mid p
\rangle / \langle p \mid \int j_0(x)d^3 x \mid p \rangle
\ee
Consider now the next order terms in $Q^2$ for the spin 1/2 target. The
situation here is more complicated, because of presence of static magnetic
moment. For definiteness consider the most interesting case of virtual
photon-proton scattering. It is instructive to calculate the contribution of
intermediate proton state to forward virtual photon-proton scattering
amplitude in general case, $\nu\not=0$. The calculation gives:
\be
f_L(\nu,Q^2)_{proton} = -\alpha Q^2m \Biggl [ \frac{1}{4 m^4}F_M^2 (Q^2) +
\frac{1}{\nu^2 - Q^4/4}G^2_E(Q^2)\Biggr ],
\ee
where $F_M (Q^2)$  is the Pauli magnetic formfactor,
\be
F_M (Q^2) = \frac{1}{1+Q^2/4m^2} \Biggl [G_M(Q^2) - G_E(Q^2)\Biggr ].
\ee
$G_E(Q^2)$ and $G_M(Q^2)$ are Sachs proton electric and magnetic
formfactors.  At $\nu=0$  and $Q^2 \to 0$ (10) evidently reduces to (7).
From (10) is clear, that the case of $\nu=0$ is a special one: at
$\nu=Const\not=0$  and $Q^2 \to 0~f_L(\nu,Q^2)\sim Q^2$ and vanishes, as it
should be for longitudinal photon scattering amplitude.

Let us now put $\nu=0$. The contribution of the excited intermediate
state to $f_L(0,Q^2)$ may be estimated as
\be
f_L(0,Q^2)_{excit.} \sim \frac{\alpha}{m}~\frac{Q^2}{M^2-m^2+Q^2}
G^2_{pN^*}(Q^2)
\ee
for spin $1/2$ state $N^*$, where $G_{PN^*}(Q^2)$ is the $p \to N^*$
transition formfactor. (The factor $Q^2$ comes from gauge invariance). For
higher spins the additional factor of $(Q^2)^n$  appears in the numerator.
Since the contributions of excited states are proportional to $Q^2$ the
constant, independent on $Q^2$  term in the expansion of $f_L(0,Q^2)$ at
small $Q^2$ arises entirely from the proton intermediate state and can be
found from (10). So, generally, for spin $1/2$  target we get up to terms of
$O(Q^2)$
\be
f_L(0,Q^2)_{spin~1/2} = 4Z^2\frac{\alpha m}{Q^2}\Biggl [ 1 - \frac{Q^2}{2m^2}
\mu_a  - \frac{1}{3}\langle r^2_E \rangle Q^2 \Biggr ],
\ee
where $\mu_a$ is the anomalous magnetic moment.
In deriving (13) the connection of Sachs and Pauli formfactors were
exploited
\be
G_E(Q^2) = F_E(Q^2) - \frac{Q^2}{4m^2}F_M(Q^2),
\ee
\be
\langle r^2_E \rangle = -6 \frac{dF_E(Q^2)}{dQ^2} \mid_{Q^2=0}.
\ee
As can be seen from (10), (14)  (as well as from (4)) $f_L(0,Q^2)$ is zero
up to terms $O(Q^2)$ for neutral particles, for example, for neutron.

Let us
estimate finally up to what $Q^2$  the proton intermediate state
contribution is dominating in longitudinal virtual photon-proton scattering
amplitude at $\nu=0$. Since the transition formfactor $G^*_{pN}(Q^2)$ in
(12) is of the same order or smaller, than $G_E(Q^2)$ (especially for highly
excited states), then the estimation (12) gives that one expect such
dominance up to $Q^2 \sim 0.5$GeV$^2$. This expectation may be checked by
calculation of the isobar intermediate state contribution to the forward
scattering amplitude of virtual  longitudinal photon on the proton at
$\nu=0$. It can be shown, that only Coulomb quadrupole  $p \to\Delta$
transition
formfactor $G^*_c(Q^2)$ contributes here. The calculation gives (in
the notation of ref.5):
$$f_L(0,Q^2)_{\Delta} = 12\alpha Mm^2 \frac{Q^2}{M^2 - m^2 + Q^2}~
\frac{1}{[(M+m)^2+Q^2][(M - m)^2 + Q^2]}\times
$$
\be
\times \Biggl( 1 + \frac{Q^2}{M^2}\Biggr)
\Biggl ( 1 + \frac{M}{m}\Biggr )^2 \Biggl ( 1 - \frac{M}{m}\Biggr )
 [ G^*_c (Q^2)]^2,
\ee
where $M$  is the isobar mass. As follows from (16) $f_L(0,Q^2)_{\Delta}$ is
much smaller than $f_L(0,Q^2)_{proton}$ (10) up to $Q^2 \sim 0.5 -1$
GeV$^2$. Therefore one may expect, that in the case of the forward scattering
amplitude of longitudinal photon on the proton at zero energy, eq.(10), may
represent the whole $f_L(0,Q^2)$ up to $Q^2 \approx 0.5$GeV$^2$.

The
physical applications of these results will be the subject of a separate
publication.

I am thankful to V.M.Belyaev for useful discussions. This work
was supported in part by INTAS Grant 93-0283.

\vspace{7mm}
\noindent
[1] W.Thirring, Phil.Mag. {\bf 41}, 1193 (1950).

\noindent
[2] N.Kroll and M.Ruderman, Phys.Rev. {\bf 93}, 233 (1954).

\noindent
[3] F.E.Low, Phys.Rev. {\bf 96}, 1428 (1954).

\noindent
[4] M.Gell-Mann and M.L.Goldberger, Phys.Rev. {\bf 96}, 1433 (1954).

\noindent
[5] H.F.Jones and M.D.Scadron, Ann.Phys. {\bf 81}, 1 (1973).
\end{document}